\documentclass[12pt]{article}

\usepackage{scicite}

\usepackage{times}
\usepackage[pdftex]{graphicx}
\graphicspath{{./pdf/}{./jpeg/}}
\DeclareGraphicsExtensions{.pdf,.jpeg,.png}

\usepackage{graphicx}

\newenvironment{sciabstract}{%
\begin{quote} \bf}
{\end{quote}}

\title{A photonic interface of chiral cavity quantum electrodynamics}

\author
{Fan Zhang,$^1$ Lingxiao Shan,$^1$ Juanjuan Ren,$^1$ Xueke Duan,$^1$ \\
Yan Li,$^{1,2}$ Tiancai Zhang,$^{2,3}$ Qihuang Gong,$^{1,2}$ Ying Gu$^{1,2\ast}$\\
\\
\normalsize{$^{1}$State Key Laboratory for Mesoscopic Physics, Collaborative Innovation Center of Quantum}\\
\normalsize{Matter, Department of Physics, Peking University, Beijing 100871, China}\\
\normalsize{$^{2}$Collaborative Innovation Center of Extreme Optics, Shanxi University,}\\
\normalsize{Taiyuan, Shanxi 030006, China}\\
\normalsize{$^{3}$State Key Laboratory of Quantum Optics and Quantum Optics Devices,}\\
\normalsize{Institute of Opto-Electronics,Shanxi University, Taiyuan 030006, China}\\
\normalsize{$^\ast$ygu@pku.edu.cn.}
}

\date{}

\begin{document}


\baselineskip24pt


\maketitle

\begin{sciabstract}

Cavity quantum electrodynamics studies light-matter interactions at single quanta level.
Chiral photon-emitter coupling  in  photonic structures is characterized as unidirectional propagation  locked by the local polarization of  light.
However, to realize the strong interaction of photon-emitter with direction-locked propagation, which will bring novel applications in nonreciprocal quantum devices, has not been reported yet.
Here, we propose the coupled photonic crystal and metallic nanoparticle structure,
where through strong local field with high helicity,
the rate of circularly polarized photons emitting into photonic crystal waveguide is one order larger than that without the nanoparticle
and the linewidth of Rabi splitting spectra is about one-tenth of that with only the nanoparticle, both with $\sim$ 95$\%$ photons propagating unidirectionally, which can be utilized in directional quantum light sources.

\end{sciabstract}



Cavity quantum electrodynamics (CQED) studies  the light-matter  interaction at a single quanta level \cite{CQED1,CQED3}. With confined electromagnetic fields,  optical mode volume and mode density determine the behavior of photon-emitter coupling.
In traditional CQED,  by compressing the optical mode volume into the region of several hundreds of micrometers, weak and strong couplings have been achieved \cite{strong_weak1}.
In weak coupling regime, the spontaneous emission of the emitter can be enhanced or suppressed by the cavity modes, while in strong coupling regime, vacuum Rabi splitting and photon blockade appear through periodically exchanging energy between the emitter and cavity photons \cite{strong_weak2,strong_weak3}.
Recently, through reducing the optical mode volume into micro/nano scale, the CQED has achieved  great success in photonic crystals \cite{CQED_PC1,CQED_PC2}, plasmonic nanocavities \cite{CQED_MNP1,CQED_MNP4,CQED_MNP5,CQED_MNP6},  whispering guided resonantors \cite{WGM1}, and various hybrid photonic systems \cite{CQED_hybrid1,CQED_hybrid2,CQED_hybrid3,CQED_hybrid4},  which paves a way to nanolaser, on-chip quantum devices, and scalable quantum networks \cite{internet}.

Besides the ultrasmall mode volume, transversely confined light in the photonic structures induces the local spin of light, whose handedness has a one-to-one relation with respect to the propagation direction of the optical field  enforced by time-reversal symmetry  \cite{fiber2,PC1,Gong2018,interference}.
If one puts a circularly polarized emitter in these structures, the propagation direction of the emitted photons will be locked by the handedness of local spin, so-called the chiral photon-emitter coupling \cite{direction_soc2,PLodahl2017}.
Hence, the propagation isolation of photons is able to avoid the signal disturbance and improve transmission efficiency, which can be utilized in  some nonreciprocal quantum information components, e.g., chiral entanglement \cite{entanglement1,entanglement2}, quantum gates \cite{PC4,gate1}, switchings \cite{switching1}, isolators \cite{isolator}, and circulators \cite{WGM2}.

\begin{figure}
  \centering
  \includegraphics[width=12cm]{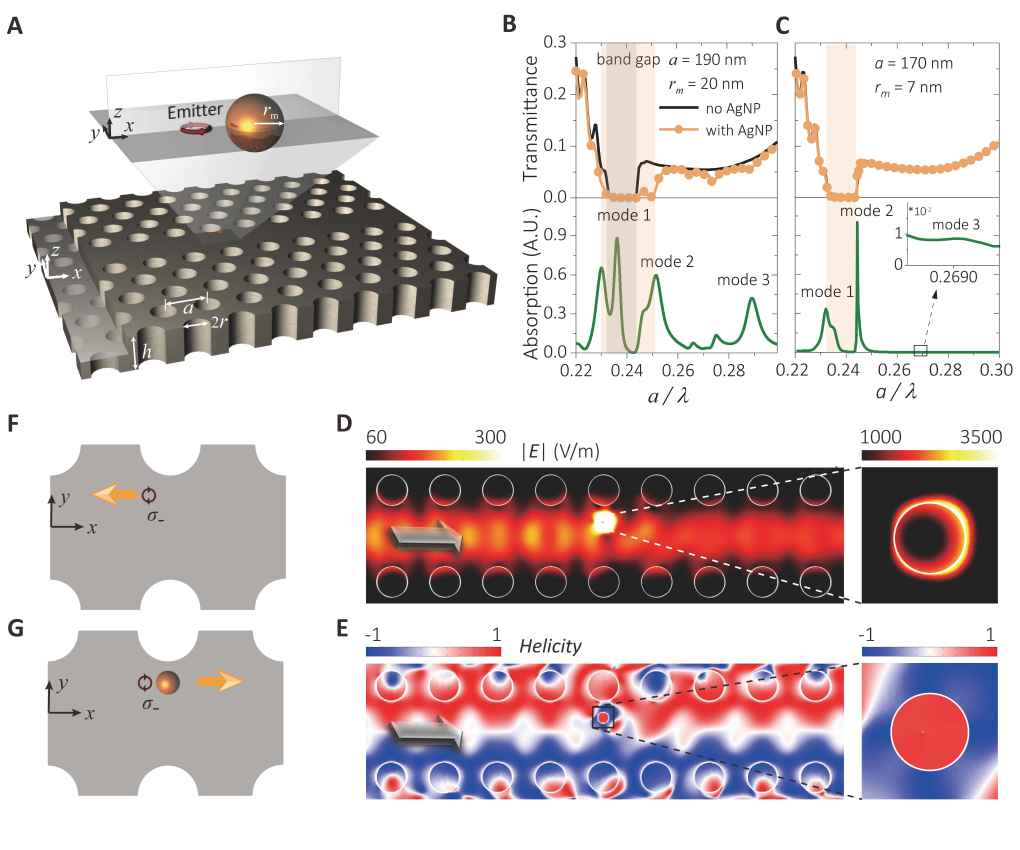}\\
  \caption{ \textbf{Optical modes of coupled W1 PC and AgNP structure.}
  (A) Schematic diagram of coupled system with $h=0.84a$ and $r=0.29a$. Transmittance spectra of PC and the absorption spectra of the AgNP embedded in the center layer of PC for (B) $a=190$ nm and $r_m=20$ nm and (C) $a=170$ nm and $r_m=7$ nm.The center of AgNP is located at the $xy$-plane with $y=50$ nm.
  Compared (B) with (C), large AgNP leads to the widening of the photonic band gap. Modes 1 and 3 are the dipolar and quadrupolar modes of the AgNP in the PC, while mode 2 of the AgNP raising at the band edge of the PC is induced by their hybridization. (D) The electric field and (E) its helicity distributions of $xy$-plane in the PC with the existence of the AgNP, where the arrows depict the propagation direction of the light. The insets show the details near the AgNP within the area of $80$ nm $\times ~80$ nm. (F) Left and  (G) right propagation of emitted photons from the $\sigma_-$ emitter  without and with the AgNP along the waveguide of PC. Opposite propagation direction comes from the coupling of spin and orbit angular momentum.
  }\label{fig1}
\end{figure}

However, in the coupling between the circularly polarized emitter and photons, to simultaneously realize the strong interaction of photon-emitter and  propagation direction lock of photons, which will bring novel applications in nonreciprocal quantum devices, has not been reported yet.
Here, by combining the advantages of the CQED and chiral coupling,
we propose the coupled structure consisting of W1 PC and Ag nanoparticle (AgNP) (Fig.~\ref{fig1}A), where high quality nanocavities are formed to guarantee an achievement of both weak and strong couplings.
Through strong local field with high helicity in nanocavities,  coupling strength of photon-emitter can be greatly enhanced, accompanied by almost unidirectional propagation of emitted photons.
In weak coupling regime, total decay rate reaches over 4500$\gamma_0$  ($\gamma_0$ is the spontaneous emission rate in vacuum), among which the rate of photon emission into the PC waveguide  is 148$\gamma_0$,  which is one order larger than that with only W1 PC \cite{PC1,reviewPC}.
While for the strong coupling, by using a low loss  band-edge mode, the linewidth of fluorescence spectra of Rabi splitting  is about one-tenth of that with only the AgNP.
For both cases, $\sim$ 95$\%$ photons propagate unidirectionally.
Moreover, though the mode design, the  coupled structure is capable of routing different wavelength photons into opposite propagation directions.
So this work bridges the fields between the CQED and the chiral coupling in photonic structures.


The electric field in the waveguide of W1 PC has non-zero spin (measured by helicity) to support single directional propagation of photons with high coupling efficiency \cite{PC3,unidirection1,coupling-eff}, but its intensity is generally very low in the position with high helicity (Fig. S1, D and E) \cite{PC4}.
So it is difficult to  arrive the strengthened photon-emitter coupling, for example, the decay rate of circularly polarized emitter in the W1 PC is only 10 times of $\gamma_0$ \cite{PC1}.
To overcome this problem, the AgNP, which has achieved success in the Purcell enhancement and vacuum Rabi splitting due to the ultrasmall mode volume \cite{CQED_MNP1,CQED_MNP4,CQED_MNP5}, is used as a  nanocavity to enhance the interaction here.
Furthermore, only depending on the individual nanoparticle structure, the propagation direction lock of photons can not be attained due to the lack of guided channel.
So combining the advantages of two structures, i.e., the high local helicity in the waveguide of  PC and the strong localized field of the AgNP, we propose the coupled structure of W1 PC and AgNP  to demonstrate  a strengthened photon-emitter coupling  with unidirectional photon propagation.

After the mode hybridization of the PC and AgNP, three optical modes arise, i.e., modes 1 and 3 are the dipolar and quadrupolar modes of AgNP, and mode 2 is the band edge mode  appearing at the edge of photonic band (Fig. 1, B and C).
Here W1 PC is chosen with the thickness $h=0.84a$, the hole radius $r=0.29a$, and refractive index $n=3.45$, where $a$ is the lattice constant, and the computations are performed by the commercial COMSOL Multiphysics and FDTD softwares (Fig. S1, B and C).
When the radius $r_m$ of AgNP is 20 nm and $a=190$ nm, there is a widening of band gap of PC due to a strong mode coupling between two structures (Fig. 1B).
In contrast, for  $r_m=7 $ nm and $a=170$ nm, the size of  AgNP is too small to affect the band of the PC (Fig. 1C).
More details of mode coupling are shown in Fig. S2.
After the parameter optimization, mode 3 in Fig. 1B,  which is situated at the guided band of the PC, acts as a nanocavity to provide large Purcell enhancement and effective guidance of photons.
While mode 2 in Fig. 1C is suitable to realize strong coupling owing to its extremely narrow linewidth.

The emitted photons in chiral photon-emitter coupling have an propagation direction lock
along the waveguide of the PC  depending on the handedness of local spin where the emitter is.
If without the AgNP,
the electric field in the PC has a symmetrical distribution, while its helicity (the helicity of $z$ direction is defined as $\frac{{2\rm Im}[E_x E_y^*]}{|E_x|^2+|E_y|^2}$ \cite{MNP1}) distribution is anti-symmetrical (Fig. S1E)  \cite{PC3}.
If we put a $\sigma_-$ emitter into the upside of the channel,
the emitted photons will excite the guided mode with right-handed spin and will be locked to the left direction (Fig. 1F).
Then, consider the hybrid PC and AgNP system. If we  take the mode 3 in Fig. 1B as an example,
 the electric field and its helicity distributions in the waveguide of the PC are almost the same as those without the AgNP  (Fig. 1, D and E) except the region around the AgNP.
Namely, around the AgNP, there is a very strong local field  (the inset of Fig. 1D) and its local helicity (the inset of Fig. 1E) has an opposite sign with that without the AgNP (Fig. S4).
According to the same principle, if one puts a $\sigma_-$ emitter into the near field region of AgNP,  the emitted photons will spread to the right direction (Fig. 1G).


\begin{figure}
  \centering
  \includegraphics[width=12cm]{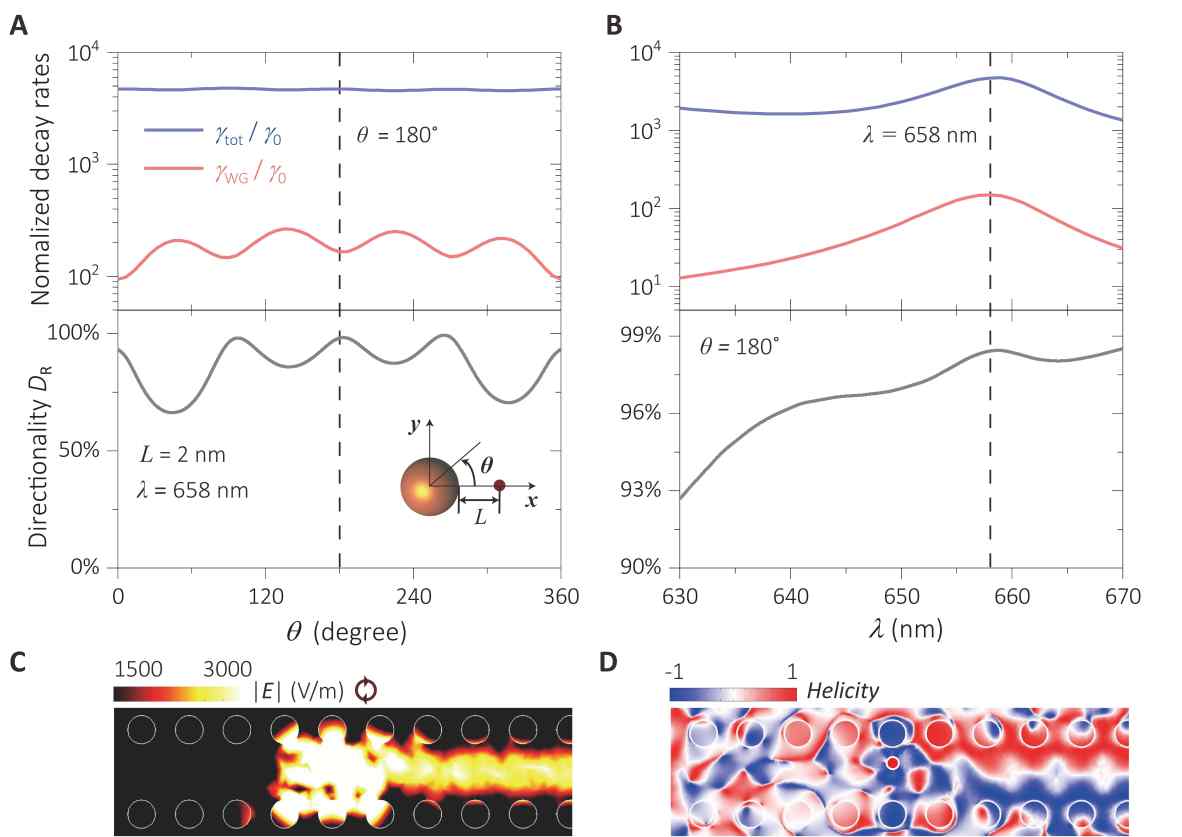}\\
  \caption{ \textbf{Nanoscale Purcell enhancement and unidirectional propagation with chiral coupling.}  Normalized decay rates $\gamma_{\rm tot} / \gamma_0$ and $\gamma_{\rm WG} / \gamma_0$ and the directionality $D_{\rm R}$ as a function of (A) $\theta$ and (B) $\lambda$. Here we use the quadrupolar mode of the AgNP (mode 3 in Fig. \ref{fig1}B) with the resonance wavelenghth of $\lambda=658$ nm to realize the chiral  coupling, and $a=190$ nm, $r_m=20$ nm, and $L=2$ nm. In (A), the maxima of $\gamma_{\rm WG} / \gamma_0$ generally correspond to the minima of $D_{\rm R}$ and vise versa, while $\gamma_{\rm tot} / \gamma_0$ keeps in a high range of 4600$\sim$4800.
In order to balance the Purcell enhancement and the directional propagation, $\theta=180^{\circ}$ is chosen in (B), where $\gamma_{\rm tot} / \gamma_0 = 4700$ and $\gamma_{WG} / \gamma_0 = 148$, and $98.4\%$ of guided part propagates along one direction due to the SOC.
The distributions of (C) the electric field and (D) its helicity when a $\sigma_-$ emitter excites the quadrupolar mode of the AgNP.
}\label{fig2}
\end{figure}

Then, Purcell enhancement of chiral coupling at the nanoscale is demonstrated.
A $\sigma_-$ emitter is put into the near field region of the resonant AgNP (mode 3 in Fig. 1B) with the distance of $L=2$ nm.
When it circles around the AgNP in the $xy$ plane, i.e. $\theta$ is changed from $0^{\circ}$ to $360^{\circ}$, Purcell factor (here the normalized total decay rate $\gamma_{\rm tot} / \gamma_0$ ) remains very high values of $4500\sim 4800$  (Fig. 2A), among which the guided part $\gamma_{\rm WG} / \gamma_0$ is $95\sim 264$ with the extreme values corresponding to the electric field  maxima of  the AgNP (the inset of Fig. 1D).
It is noted that the guided part $\gamma_{\rm WG} / \gamma_0$ alone is one order larger than the $\gamma_{\rm tot} / \gamma_0$ ($\sim 10$)  of single W1 PC (Fig. S5).
There is also a correspondence between the maximum of directionality $D_{\rm R}$ and the minimum of decay rates, where $D_{\rm R/L}=\frac{W_{\rm R/L}}{W_{\rm R}+W_{\rm L}}$ and $W_{\rm R/L}$ is the energy power from the emitter into the right or left end of the channel (Fig. S6).
This correspondence can be explained as follows.
If the $\sigma_-$ emitter is placed at the positions with the higher helicity, more emitted photons couple to the eigenstate mode with the same spin, leading to the maxima in $D_{\rm R}$ curve.
In contrast, if the emitter is located at the lower helicity region,  such as for $\theta=45^\circ$, the directionality $D_{\rm R}$ is small, but two channels (to  left and right directions) can transmit more  photons (Fig. S9), so the maxima of $\gamma_{\rm WG} / \gamma_0$ appear.
By comparing these curves with the insets of Fig. 1, D and E, it is proved that strong local field leads to a large Purcell enhancement and high helicity results in a good directionality.


To balance the inconsistence between the decay rate and directionality,  in Fig. 2B, the parameter of $\theta=180^{\circ}$ is chosen.
At the resonance wavelength of $\lambda=658$ nm, both the Purcell enhancement in guided part and the directionality reach the highest values.
Especially, in the spectral range of 655 nm $\sim$ 661 nm, $\gamma_{\rm tot} / \gamma_0$  and  $\gamma_{\rm WG} / \gamma_0$ can reach 4200 and 110 as well as $98\%$ of the guided photons  propagates unidirectionally.
Fig. 2, C and D depicts the electric field  and local helicity distributions when the electromagnetic mode is excited by the $\sigma_-$ emitter with $\lambda=658$ nm.
As mentioned above,  the existence of the AgNP changes the local helicity of the nanoarea where the emitter is.  So  the propagation direction is opposite to that if without a AgNP.
Moreover, if now we put a $\sigma_+$ emitter near the AgNP, the  photons will be launched to the opposite direction according to the symmetry (Fig. S10).

Furthermore, the coupled structure is also capable of separating the different wavelength photons into opposite propagation directions.
For example, in present coupled system, if the AgNP is substituted by a silver nanoblock with the length of 25 nm and the height of 10 nm, its optical modes appear at the wavelength of $706$ nm and $639$ nm respectively (Fig. S11A).
As shown in Fig. S11, E and F, if putting a $\sigma_-$ emitter with $\lambda=706$ nm at the corner of the block, the guided part of emitted photons with $\gamma_{\rm WG} / \gamma_0=389$ is steered to the left direction with the directionality of $D_L=73\%$, while for the emitter with $\lambda=639$ nm, the guided photons with $\gamma_{\rm WG} / \gamma_0= 76$ are selected to the right directions with $D_R=89\%$ (Fig. S11, B and G).
The reason is that, for different resonant wavelengths, the local helicity at the corner of the block has the opposite sign, i.e., more than zero or less than zero (Fig. S11, C and D). So the photons with the same spin will propagate to different directions, which may be used in on-chip routing single photons source.


\begin{figure}
  \centering
  \includegraphics[width=12cm]{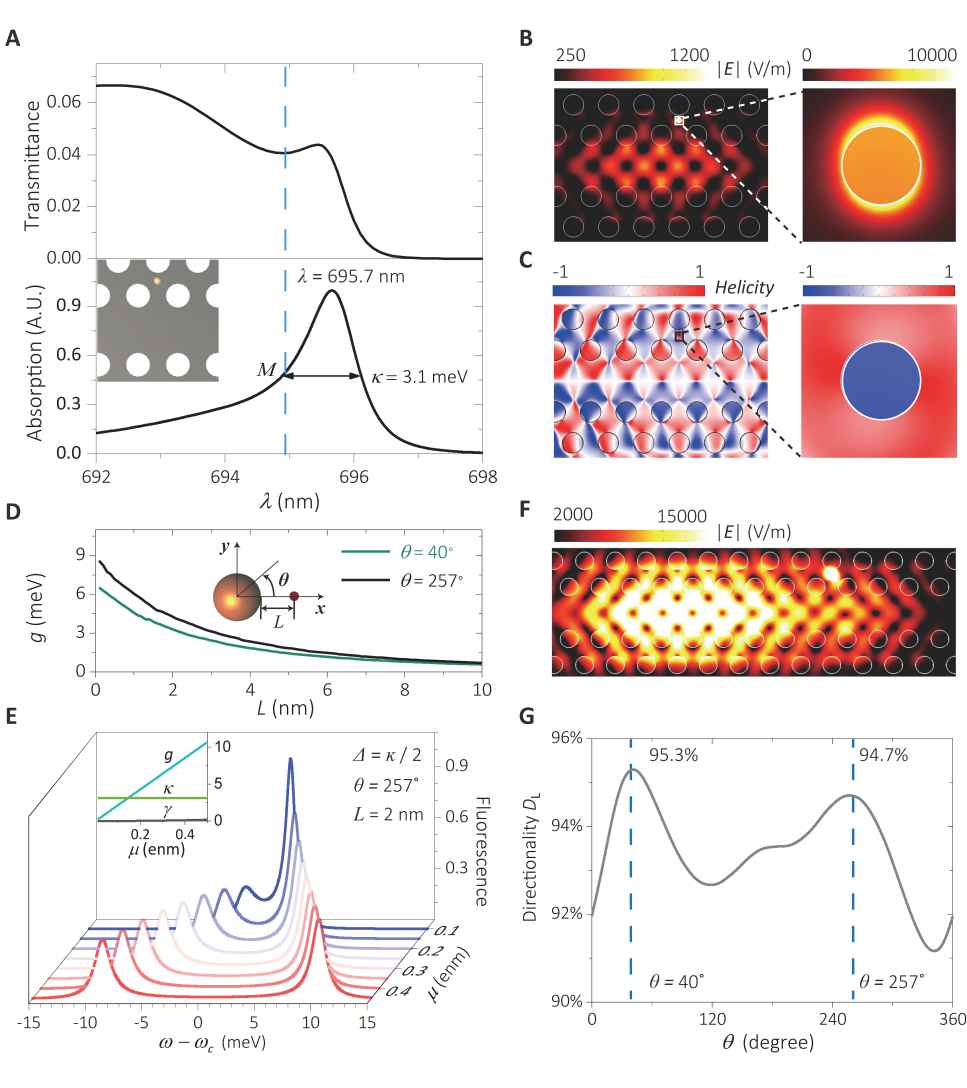}\\
  \caption{ \textbf{Rabi splitting of fluorescence spectra and uni-directional propagation with chiral coupling.} (A) Transmittance spectra of PC and absorption spectra of the AgNP of the band edge mode (mode 2 in Fig. 1C). Here $a=170$ nm and $r_m=7$ nm. The mode at $\lambda=695.7$ nm, raised by the mode hybridization of PC and AgNP, has a very narrow linewidth of $\kappa=3.1$ meV. (B) The electric field and (C) its helicity distributions of the band edge mode with the insets of the region of $30$ nm $\times ~30$ nm. (D) Coupling coefficient $g$ between the $\sigma_-$ emitter and photon with varying the distance $L$ for $\theta=40^{\circ}$ and $\theta=257^{\circ}$. (E) Fluorescence spectra of the quantum emitter as a function of $\omega-\omega_c$ with varying dipole moment $\mu$. Here $\omega_c=2\pi c/\lambda$. The inset shows the parameters $g,~\gamma,~\kappa$ in dependence on $\mu$.
The asymmetry of fluorescence spectra comes from the nonzero detuning $\Delta$ between the emitter and the nanocavity.
  (F) The electric field distribution for $\theta=257^{\circ}$ when a $\sigma_-$ emitter excites the band edge mode. (G) Directionality $D_{\rm L}$ of guided photons with varying $\theta$.
  In (E) to (G),  the detuning $\Delta=\kappa/2$.
  }\label{fig3}
\end{figure}

Next, the vacuum Rabi splitting of energy levels in the chiral coupling is demonstrated.
We put the AgNP near the waveguide of the PC (inset of Fig. 3A) and demonstrate that the appearance of the band-edge mode is independent of the position of the AgNP (Fig. S12).
In Fig. 3A, the mode (mode 2 in Fig. 1C) appearing at the edge of photonic band with $\lambda=695.7$ nm has a very narrow linewidth of $\kappa=3.1$ meV, which is only one-tenth of the dipole mode of an AgNP in homogenous medium (Fig. S3).
So for $\lambda=695.7$ nm, $\Delta \lambda=1.2$ nm and its quality factor $Q$ is 580.
Corresponding to the mode excitation,  there is an electric field enhancement inside the waveguide of the PC, but the local  field around the AgNP is two orders stronger than that surrounding it in the waveguide (Fig. 3B).
Then, taking $\theta=40^{\circ}$ and $\theta=257^{\circ}$ as examples, with varying the distance $L$, coupling coefficients $g$ between a $\sigma_-$ emitter with the dipole moment  $\mu=0.2$ enm and the nanocavity are shown in Fig. 3D.
For both cases, $g$ decreases exponentially as an increment of $L$.
Especially, for $L=2$ nm and $\theta=257^{\circ}$, $g=4.28$ meV,  $\kappa=3.1$ meV, and the decay rate $\gamma=0.03$ meV.  Thereby,  the strong coupling condition $g>(\kappa,\gamma)$ can be easily satisfied \cite{Quantum}.

By using the Python toolbox, the fluorescence spectra of a $\sigma_-$ emitter coupled with the band edge mode are obtained.
Considering that the directionality $D_{\rm L}$ at $\theta=257^{\circ}$ is $94.7\%$ (Fig. 3G) and the point $M$ with the frequency detuning $\Delta=\kappa/2$ (between the emitter and cavity mode)  is within the guided band of the PC which is benefit to the photons transmittance (Fig. 3A), we choose the parameters of $\theta=257^{\circ}$, $L=2$ nm, and  $\Delta=\kappa/2$.
It is found that the Rabi splitting in fluoresence spectra starts to appear at $\mu=0.1$ enm and becomes larger with its increment; while for $\mu=0.2$ enm, there is an apparent  energy exchange between the emitter and cavity photons (Fig. 3E).
Besides, owing to the existence of detuning $\Delta$, the symmetry of  fluorescence spectra is broken.
The energy splitting and the spectral linewidth are roughly coincident with $\sqrt{4g^2+\Delta^2-\frac{(\kappa-\gamma)^2}{4}}$ and  $\frac{\kappa+\gamma}{2}$ of the dressed state theory  \cite{Quantum,Rabimode}.
The linewidth of fluorescence spectra is about one-tenth of that if only  the existence of the AgNP.
Hence, superior to the properties of individual AgNP, the narrower linewidth of the band edge mode leads to the earlier appearance  of Rabi splitting with the smaller $\mu$.

In this case, the sign of electric field helicity around the AgNP (the inset of Fig. 3C) is opposite to that in Fig. 1E and Fig. 2D.
So if now putting a $\sigma_-$ emitter into its near field region, we can see that the photons will propagate in opposite direction (Fig. 3F), i.e., along $-x$ direction.
Compared with the case of $\Delta=0$ (Fig. S13), more photons can be transmitted because  at the point $M$ the frequency of emitted photons lies in the guided band of the PC.
For $L=2$ nm and $\Delta=\kappa/2$, if letting the emitter walk a circle around the AgNP, the directionality $D_L$ reaches its maximum at $\theta=40^\circ$ and  $257^{\circ}$.
Corresponding to these maxima, $\sim 95\%$ of guided photons  propagates into one direction along the waveguide of the PC, which may be used in the nonreciprocal quantum nanophotonic devices \cite{PLodahl2017}.

Finally, we address the fabrication possibility of our scheme. Nowadays, single AgNP \cite{CQED_MNP1} and PC \cite{coupling-eff} can be fabricated by state-of-the-art nanotechnology. Besides, single emitters embedded in photonic-crystal waveguide in the experiments have also been realized through scanning tunneling microcopy \cite{reviewPC}. The main challenge to realize efficient chiral coupling is how to control the relative positions between the emitter and the AgNP precisely. This problem may be solved by using an atomic force microscopy tip to move the AgNP after the position of an emitter is fixed \cite{movable}.
To obtain circularly polarized photons with different frequency, first two circularly polarized states $|+\rangle$ and $|-\rangle$ of the atoms \cite{fiber1} or quantum dots \cite{PC4} are generated by applying a strong magnetic field in $z$ direction;
then, by selecting the frequency of the excitation laser, the emitters only emit $\sigma_-$ or $\sigma_+$ photons. Thus, it may be possible to achieve our scheme in experiment in near future.


In summary, we have established a photonic interface of chiral CQED by proposing  the coupled photonic crystal and plasmon nanoparticle structure.
We clearly state that the key element of chiral CQED is the  joint action of strong local field and its high helicity,  which provides a strengthened light-emitter coupling with good directionality of emitted photons.
Using the basic idea presented here, other kinds of combined photonic structures could be designed, such as the coupled nanowire and nanoparticle,  photonic crystal and nanocavity systems, and so on.
The results  bridge the fields of the CQED and the chiral coupling, which greatly enriches the contents of light-emitter interaction at the nanoscale.
By combining the advantages of these two fields, we  provide a possible platform  for on-chip nonreciprocal quantum light sources, quantum circuits, and scalable quantum network.

 \section*{Acknowledgement}

 We thank X. Hu and Y. Ao for helpful discussions. This work is supported by the National Key R$\&$D Program of China under Grant No. 2018YFB1107200, and by the National Natural Science Foundation of China under Grants No. 11525414 and No. 11734001.


\appendix

\section*{Material and Methods}

\subsection*{1. Computation module}

We use the commercial COMSOL multiphysics software to perform the simulations. The TE-like PC, containing $11 \times 12$ unit cells shown in Fig. S1A, is placed in the middle of a three-dimensional module with the hight of $h+1~\mu$m (Fig. S1B). To minimize boundary reflections and form an infinite space, scattering boundary condition is used to surround the module. An AgNP, whose permittivity is taken from the experimental data~\cite{Johnson1972}, is embedded in the middle plane of the waveguide of PC. By incidenting a beam of plane wave propagating along the $x$-axis, or putting an oscillating point dipole inside the PC, modes of the coupled PC and AgNP structure can be excited.

The circular polarization of the light is represented by helicity, which is defined as \cite{MNP1}
\begin{equation}\label{polarization}
  C = \frac{|E_{\rm LCP}|^2 - |E_{\rm RCP}|^2}{|E_{\rm LCP}|^2 + |E_{\rm RCP}|^2},
\end{equation}
where $E_{\rm LCP/RCP}$ is the left- or right-component of the electric field which is related to the basis. If we choose the basis of $\frac{\hat{x}+i\hat{y}}{\sqrt{2}}$, $\frac{\hat{x}-i\hat{y}}{\sqrt{2}}$, and $\hat{z}$, the helicity in the $z$-direction can be written as $C(z)=\frac{2Im[E_xE_y^*]}{|E_x|^2+|E_y|^2}$. In the same way, we can derive $C(x)=\frac{2Im[E_yE_z^*]}{|E_y|^2+|E_z|^2}$ and $C(y)=\frac{2Im[E_zE_x^*]}{|E_z|^2+|E_x|^2}$. From Eq. (\ref{polarization}), $C=1$ represents the left-polarized light and $C=-1$ is the right-polarized light.
In the COMSOL module, the left- and right-handed polarized plane waves along z-axis are given as $\vec{E}_{\pm}=E_x\hat{x}+E_y\hat{y}=\frac{E_0\hat{x} + E_0e^{\mp \pi/2}\hat{y}}{\sqrt{2}}$, where $E_0$ is the amplitude of the electric field and $E_y$ is $\pm \pi/2$ out of phase with $E_x$.
Similarly, the left- and right-handed polarized emitters are set as oscillating point dipoles for $\sigma_{\pm}=\frac{\mu \hat{x} + \mu e^{\mp \pi/2}\hat{y}}{\sqrt{2}}$, where $\mu$ is the magnitude of the dipole moment.
The transmittance spectra of the PC is defined as $I_t/I_0$, where $I_t$ and $I_0$ are the light intensity in the incident and exit surfaces of the PC. In this text, the light is incident on the left end of the PC and the right end is used as receiving surface to collect photons.

\subsection*{2. Photonic band diagram of photonic crystal}

We compute the photonic band diagram of the PC by a finite-difference time-domain (FDTD) method with commercial software (Lumerical) (Fig. S1C). To reduce the computation process, a two-dimensional module, where the thickness of the PC is represented by a modified refractive index $n_{\rm eff}$, is performed. Through boundary mode analysis in the COMSOL module, we obtain $n_{\rm eff}=2.95$. Compared with the transmittance spectra in Fig. 1, B and C, the band gap in Fig. S1C is coincident with that in the transmittance spectra obtained by COMSOL software.

\subsection*{3. Computation of the coupling coefficients}

Three physical processes are included in the CQED systems: the coupling between the cavity photons and the emitter, the decay of the emitter, and the cavity loss, whose coefficients are labeled as $g$, $\gamma$, and $\kappa$, respectively.
According to their relations, two typical regimes exist, i.e., weak coupling for  $g << \gamma,~\kappa$ and strong coupling for $g >> \gamma,~\kappa$ \cite{strong_weak1}.

\subsubsection*{Weak coupling}

In weak coupling regime, the emitter decays through three channels: guiding along the waveguide, radiating into far field and nonradiative loss. Thus the total decay rate $\gamma_{\rm tot}$ is equal to the sum of these decay rates, i.e., $\gamma_{\rm tot}=\gamma_{\rm rad}+\gamma_{\rm nr}=\gamma_{\rm WG}+\gamma_{\rm free}+\gamma_{\rm nr}$, where $\gamma_{\rm free}$ is the decay rate to the free space.
The total normalized decay rate can be obtained from $\gamma_{\rm tot}/\gamma_0=W_{\rm tot}/W_0$ \cite{Lukas-book}, where $W_{\rm tot}$ and $W_0$ are the total emitted energy power of an emitter in the coupled system and in vacuum, respectively. The energy power of the emitter in the COMSOL module is given by the surface integration of a nanosphere containing the emitter over the energy flows, which can be expressed by $W_{\rm tot}=\int\int_{\Sigma}\vec{S}\bullet d\Sigma$ and $\vec{S}$ is the Poynting vector on the nanosphere \cite{CQED_hybrid3}.
Similarly, the guided part along the waveguide is calculated by $\gamma_{\rm WG}/\gamma_0=W_{\rm WG}/W_0$, where $W_{\rm WG}$ is the energy power in the receiving surfaces--right and left surfaces of the module.

\subsubsection*{Strong coupling}

The dynamics of an emitter-nanocavity coupled system in strong coupling regime can be represented by JC-model, where the Hamiltonian and dynamical equation are given by
\begin{equation}\label{H}
  H=\omega_e\sigma^+\sigma+\omega_ca^+a+g(\sigma^+a+a^+\sigma),
\end{equation}
and
\begin{equation}\label{rho}
  \dot{\rho} = i[\rho,H]/\hbar + \frac{\gamma}{2}(2\sigma \rho \sigma^+ -\{\sigma^+ \sigma,\rho\}) + \frac{\kappa}{2}(2a\rho a^+ - \{a^+a,\rho\}),
\end{equation}
where $\omega_e$ and $\omega_c$ are the frequencies of the emitter and the nanocavity, $\sigma$ ($\sigma^+$) is lowering (raising) operator of the emitter and $a$ ($a^+$) is bosonic annihilation (creation) operator of the AgNP, respectively. The physical process in this system is that the emitter and the nanocavity exchange energy with coupling coefficient $g$, and simultaneously, photons also lose because of the atomic decay and Ohmic loss of the cavity, with coefficients labeled as $\gamma$ and $\kappa$. In the following we show how to compute these coefficients in the COMSOL software.
\\

\underline{a. Coupling coefficient $g$}

The AgNP is excited by the light propagating along the waveguide of the PC or by the background electric field (the case for the AgNP in the homogeneous medium).
Following the equations and method in Refs. \cite{Sowik2013,CQED_hybrid4}, $\hbar g = \mu E_{\rm s}$ where $E_{\rm s}$ is the electric field of AgNP corresponding to a single excitation and can be written as
\begin{equation}\label{1}
  E_{\rm s}=\widetilde{E}/\sqrt{\frac{W}{\hbar \omega_c}},
\end{equation}
where $\widetilde{E}$ is the excited electric field of the nanocavity, $\sqrt{\frac{W}{\hbar \omega_c}}$ symbols the number of photons with energy of $\hbar \omega_c$, and $W$ is the total energy of the cavity mode which is calculated by energy density integration in the whole space \cite{CQED_hybrid4}:
\begin{equation}
  W=\frac{1}{2}\int{\frac{\partial}{\partial \omega}[\omega Re[\varepsilon(\omega)]]|_{\omega=\omega_c}|\widetilde{E}|^2dV}+\frac{1}{2}\int{\mu_0|\widetilde{H}|^2dV}.
\end{equation}
\\

\underline{b. Cavity loss $\kappa$}

The cavity loss $\kappa$ is the fullwidth at half-maximum (FWHM) of its extinction spectra, including the scattering and absorption of the cavity. Because the FWHM of scattering and absorption spectra are almost the same~\cite{Evlyukhin2007}, $\kappa$ is derived from the absorption spectrum here. In the COMSOL module, the mode of AgNP is excited by a nearby oscillating point dipole. By volume integrating the AgNP over the power density, the resistive loss $W_{\rm nr}$ is obtained. $\kappa$ of the AgNP in the homogeneous environment with refractive $n=3.45$ is shown in Fig. S3.
\\

\underline{c. Decay rate $\gamma$}

Decay rate $\gamma$ here means that the atomic decay rate to the modes other than the cavity mode, i.e., except for the nonradiative decay caused by the loss of the AgNP. Therefore, $\gamma$ is equal to the total decay rate minus the nonradiative part, i.e., $\gamma=\gamma_{\rm tot}-\gamma_{\rm nr}$, which can be obtained by $\gamma=\frac{W_{\rm tot}-W_{\rm nr}}{W_0}\gamma_0$ with $\gamma_0=\omega^3\mu^2/3\pi\epsilon_0\hbar c^3$~\cite{Quantum2}.

\subsection*{4. Fluorescence spectra of the quantum emitter}

Python toolbox is performed to derive the resonance fluorescence spectrum of the CQED system by Fourier transformation electric intensity $\langle E^-(\vec{r},t)E^+(\vec{r},t) \rangle$, with the expression of $S(\vec{r},\omega_0)=\frac{1}{\pi}Re\int d\tau \langle E^-(\vec{r},t)E^+(\vec{r},t) \rangle e^{i\omega_0 t}$ \cite{Quantum2,qutip}. By increasing the dipole moment of the emitter,  Rabi splitting appears in the fluorescence spectra (Fig. 3E and Fig. S13). When the transition frequency of the emitter is coincident with the resonant frequency of the nanocavity, the peaks of Rabi splitting are symmetry (Fig. S13). By contrast, asymmetry occurs as the emitter frequency deviates from that of the AgNP mode (Fig. 3E).

\section*{Supplementary Text}

\subsection*{1. The electric field and its helicity for the modes of the AgNP}

In the following, we analyse the electric field and its helicity for the mode of the AgNP in a homogeneous medium with the refractive index of $n=3.45$.
Figure~S4, A-B, and C-D depicts the electric field and helicity  distributions of the AgNP in the $xy$- and $yz$-plane, respectively. An AgNP with the radius of 7 nm is excited by a right-handed plane wave polarized in the $xy$-plane. The wavelength of the incident light is 625 nm, corresponding to the wavelength of the dipole mode. It is seen that the electric field in the $yz$-plane is similar to that of a linearly polarized dipole in the $y$ direction. While in the $xy$-plane, the absolute value of the electric field is almost  homogeneous surrounding the AgNP. To be mentioned, the helicity around the AgNP in the $xy$ plane has an opposite sign against that of the excited light. This phenomenon can be explained as follows. The equivalent dipole moment of the AgNP excited by a plane wave with $\vec{E}=[1, i, 0]E_0 e^{-i(k_z z-\omega t)}$ is given as
\begin{equation}\label{Emnp}
  \vec{p}_{\rm MNP}=4 \pi \varepsilon_b \varepsilon_0 r_m^3 \Gamma ({E_0 \hat{x}+iE_0 \hat{y}}),
\end{equation}
where $\Gamma=(\varepsilon_m - \varepsilon_b) / (\varepsilon_m + \frac{n^{'}+1}{n^{'}} \varepsilon_b)$ with the permittivity $\varepsilon_m$ of the AgNP and $\varepsilon_b$ of the host medium \cite{Ridolfo2010a}.
Take the dipole mode as example, namely, $n^{'}=1$, $\Gamma=(\varepsilon_m - \varepsilon_b) / (\varepsilon_m + 2 \varepsilon_b)$ and the resonant condition is $\varepsilon_m = -2 \varepsilon_b$. Under the resonant condition, $[p_x, p_y] \propto [-i, 1]$, which has the same form with that of a $\sigma_-$ emitter whose near field has $>0$ helicity in the $yz$-plane. Similarly, the local polarization of the higher-order modes of the AgNP also has the same sign as that of the dipole mode.

\subsection*{2. Optical mode coupling between the PC and the AgNP}

The transmittance spectra of the PC and the absorption spectra of the AgNP with various radius $r_m$ are shown in Fig. S2. The same as in Fig. 1, B and C, mode 1 and mode 3 correspond to the dipole and quadrupole modes of the AgNP and mode 2 is a band-edge mode. Here, lattice constant $a$ is changed to make the mode of the AgNP situating at different frequency regions of PC, i.e., guided and band gap regions. Take $r_m=7$ nm for example. As $a$ increases, the dipole mode of the AgNP moves from the band gap to the guided band of the PC. Comparing Fig. S2, A to C, it is seen that narrow and sharp band-edge modes appear when the dipole mode of the AgNP locates at the band gap of the PC (Fig. S2C).

The gray region depicts the band gap of the PC without the existence of the AgNP. It is seen that the band gap is broadened as the radius $r_m$ of the AgNP enlarges. Red shift of mode 1 and mode 3 in the absorption spectra is also clearly shown as an increment of $r_m$.

\setcounter{figure}{0}
\renewcommand{\thefigure}{S\arabic{figure}}

\clearpage
\begin{figure}
  \centering
  \includegraphics[width=15cm]{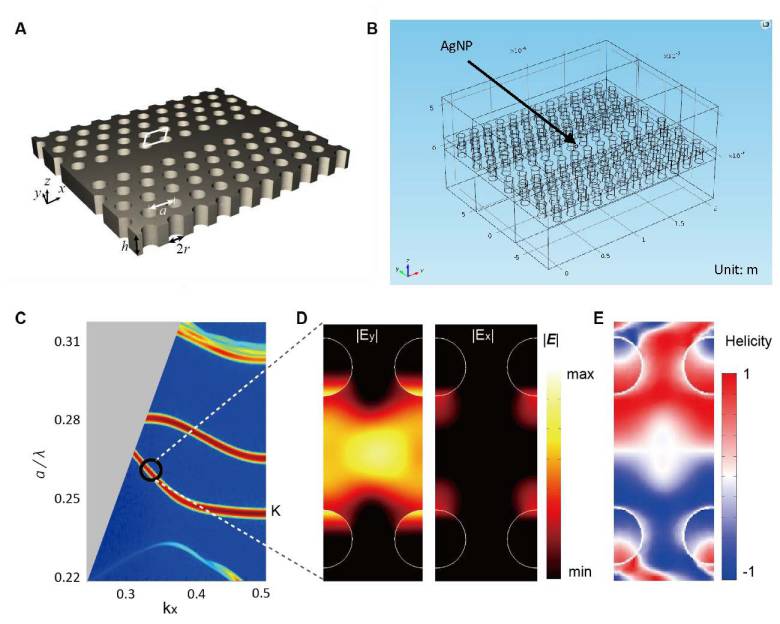}\\
  \caption{(A) Schematic diagram of the W1 PC and (B) computation module of COMSOL software. One unit cell of the PC is marked within the white lines in (A). (C) Photonic band diagram of the coupled W1 PC and AgNP structure for the period of $a=190$ nm, the hole radius of $r=0.29a$, the thickness of $d=0.84a$, and the refractive index of $n=3.45$. (D) The electric field components $|E_x|$ and $|E_y|$ and (E) helicity distributions of $z$ direction when $a/\lambda=0.26$. The electric field intensities are normalized by the maximum of $|E_y|$. }\label{figS1}
\end{figure}

\clearpage
\begin{figure}
  \centering
  \includegraphics[width=15cm]{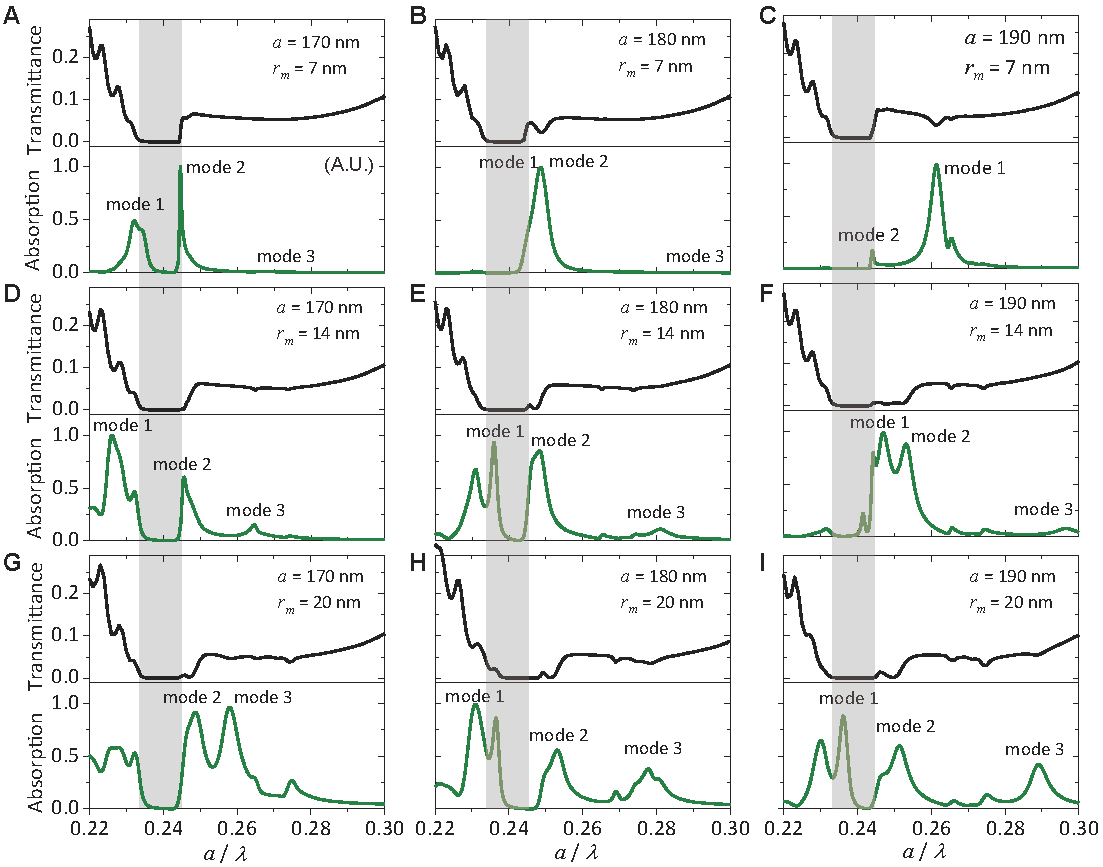}\\
  \caption{The transmittance spectra of the PC and the normalized absorption spectra of the AgNP with the radius of (A-C) $r_m=7$ nm, (D-F) $r_m=14$ nm, and (G-I) $r_m=20$ nm. The parameters for $a=170$ nm$\sim 190$ nm are computed to change the coupling between the modes of the PC and AgNP. Fig. 1, B and C in the main text corresponds to (A) and (I), respectively, and the band gap of a PC without the AgNP is shown as the gray region. Mode 1 and mode 3 are the dipole and quadrupole modes of the AgNP and mode 2 is the band-edge mode corresponding to the K point in the band diagram of Fig. S1C. By increasing the radius of the AgNP, the band-edge mode always exists and its linewidth is increased. Specifically, when $a=180$ nm and $r_m=7$ nm , mode 1 is at the band edge of the PC so modes 1 and 2 are almost overlapped.
 }\label{figS2}
\end{figure}

\clearpage
\begin{figure}
  \centering
  \includegraphics[width=12cm]{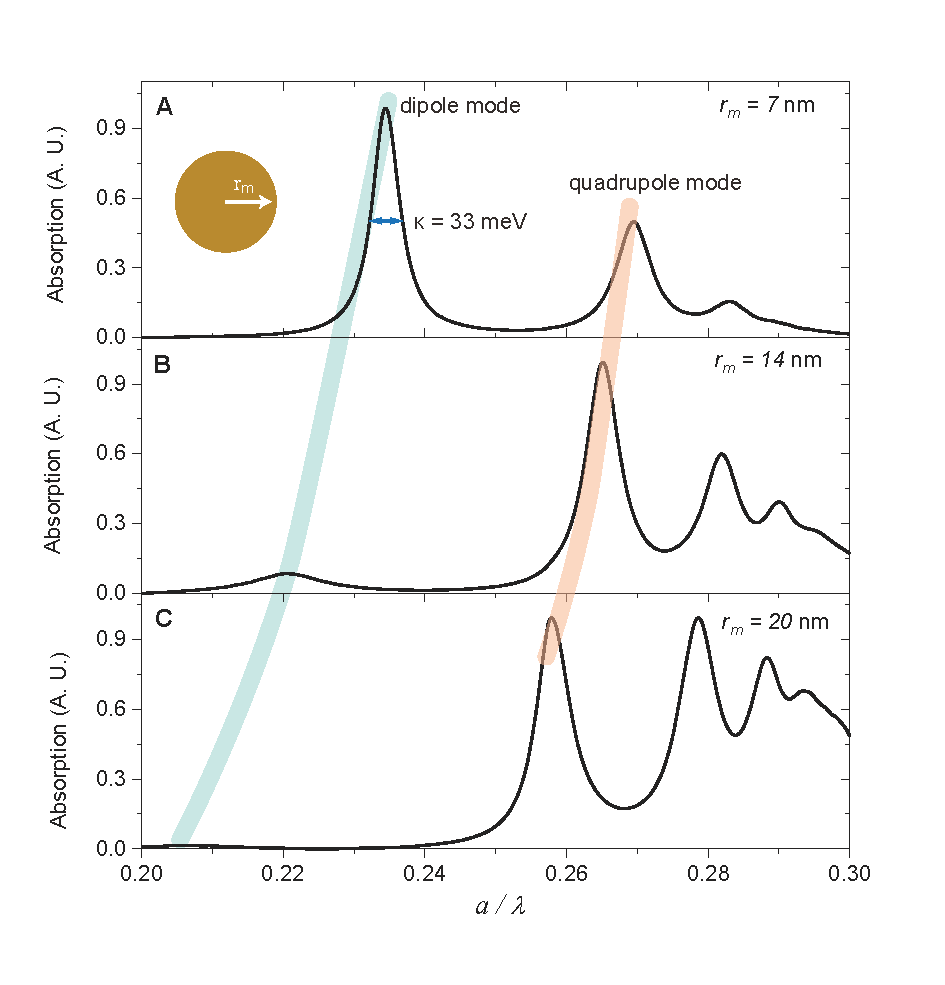}\\
  \caption{Absorption spectra of the AgNP in a homogeneous medium with refractive index of $n=3.45$, for the radius of (A) $r_m=7$ nm, (B) $r_m=14$ nm, and (C) $r_m=20$ nm, respectively. The wavelength is normalized by $a=170$ nm to compare with the case of the AgNP embedded in the PC waveguide. For the AgNP with different size, red shift in the resonance wavelength occurs as $r_m$ enlarges. }\label{figS3}
\end{figure}

\clearpage
\begin{figure}
  \centering
  \includegraphics[width=12cm]{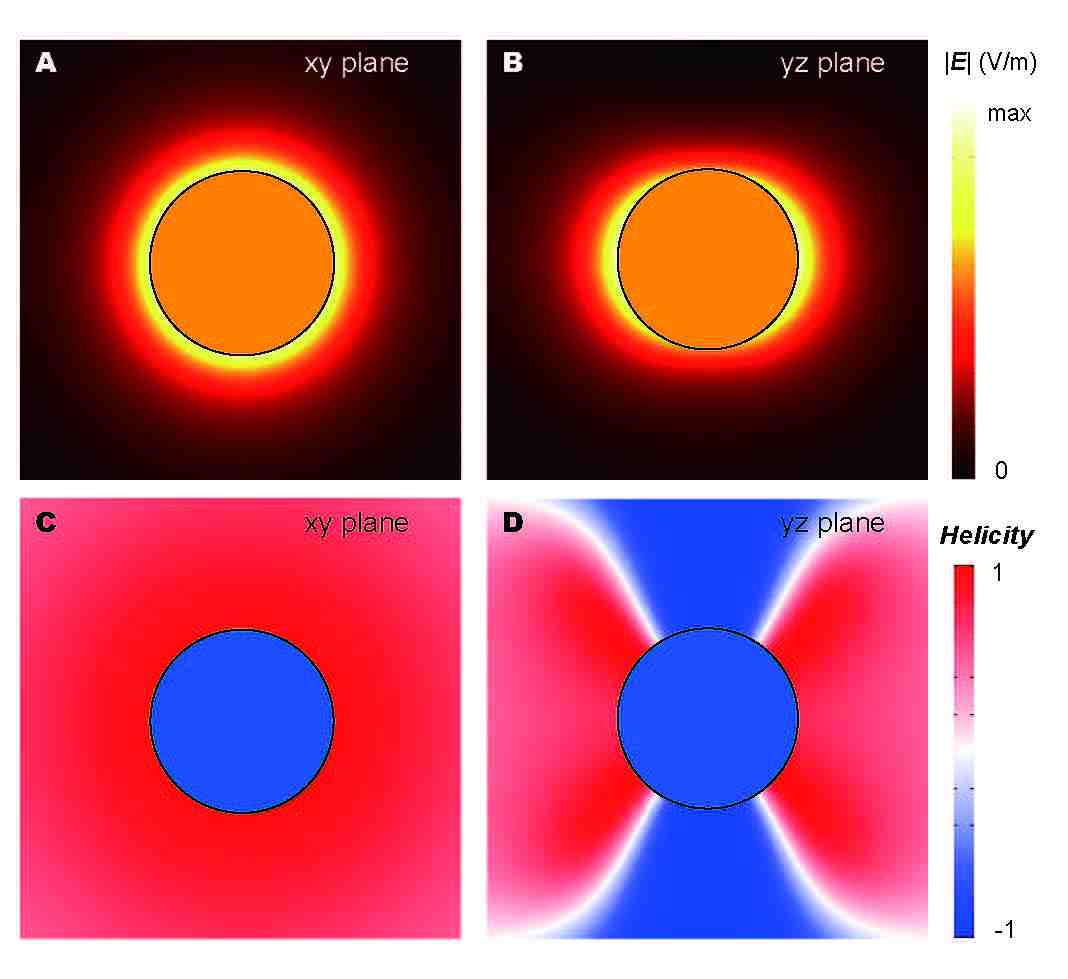}\\
  \caption{The electric field and helicity distributions of the dipole mode of an AgNP with $r_m=7$ nm in the (A and C) $xy$ and (B and D) $yz$ planes embedded in a homogeneous medium with refractive index of $n=3.45$. A right-handed polarized plane wave is used to excite the AgNP, which propagates along $z$ direction. From (C), the helicity around the AgNP in the $xy$ plane has an opposite sign against that of the excitation light.}\label{figS4}
\end{figure}

\clearpage
\begin{figure}
  \centering
  \includegraphics[width=14cm]{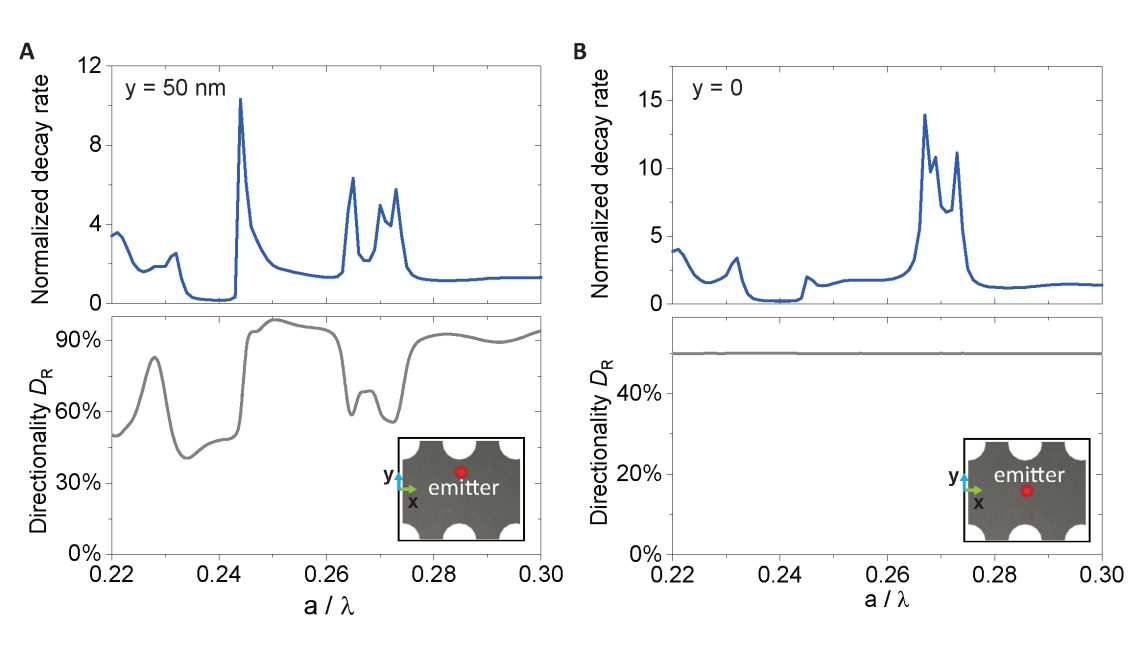}\\
  \caption{Total decay rates $\gamma_{\rm tot}$ normalized by $\gamma_0$ of a $\sigma_-$ emitter embedded in the PC waveguide for (A) $y=0$ and (B) $y=50$ nm. $\gamma_{\rm tot}/\gamma_0$ is almost less than 15 when $a/\lambda$ is within the guided mode region \cite{PC3}. The directionality $D_{\rm R}$ for the emitter placed at (C) $y=0$ and (D) $y=50$ nm. The lattice constant is $a=190$ nm. }\label{figS5}
\end{figure}

\clearpage
\begin{figure}
  \centering
  \includegraphics[width=13.5cm]{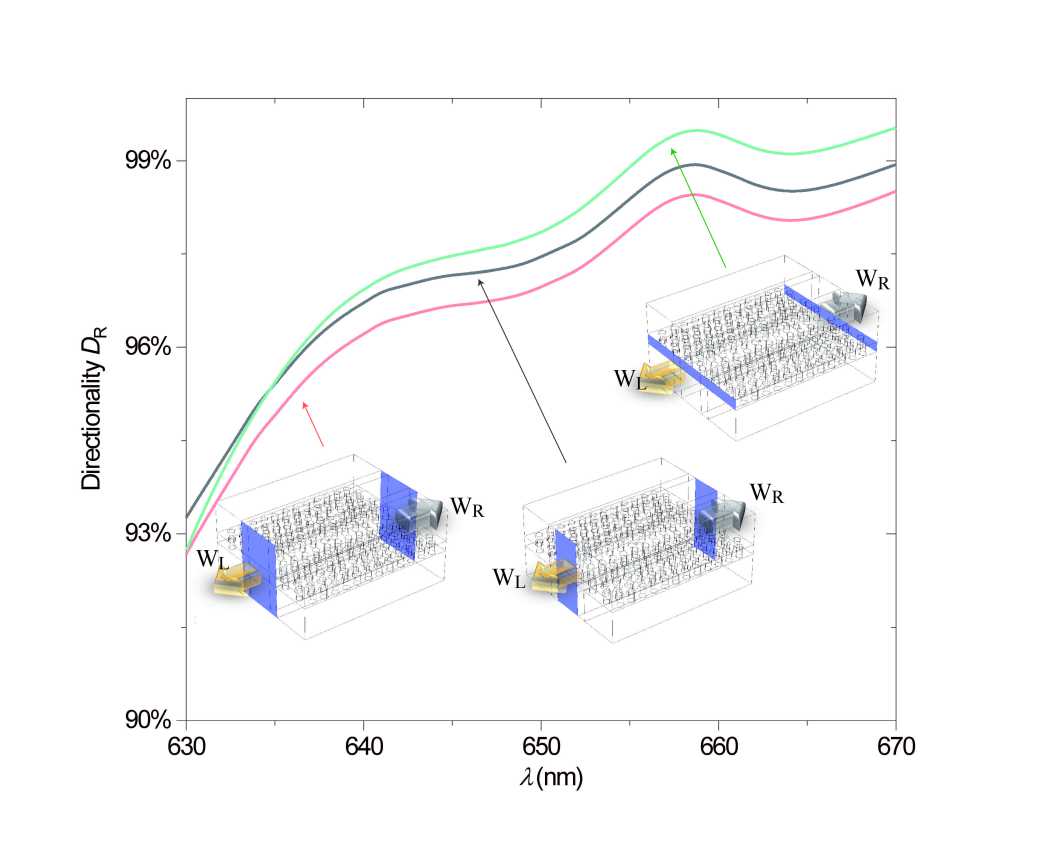}\\
  \caption{Directionality of the photons emitted from a $\sigma_-$ emitter for different receiving surfaces to collect photons shown as the insets. 
  Because the photons lose the lock of propagation direction in the free space, directionality $D_{\rm R}$ reduces a little when the surrounding environment is involved. However, the wavelength of the maxima in these three cases are not changed. Here, the emitter is set at $y=50$ nm. }\label{figS6}
\end{figure}

\clearpage
\begin{figure}
  \centering
  \includegraphics[width=13.5cm]{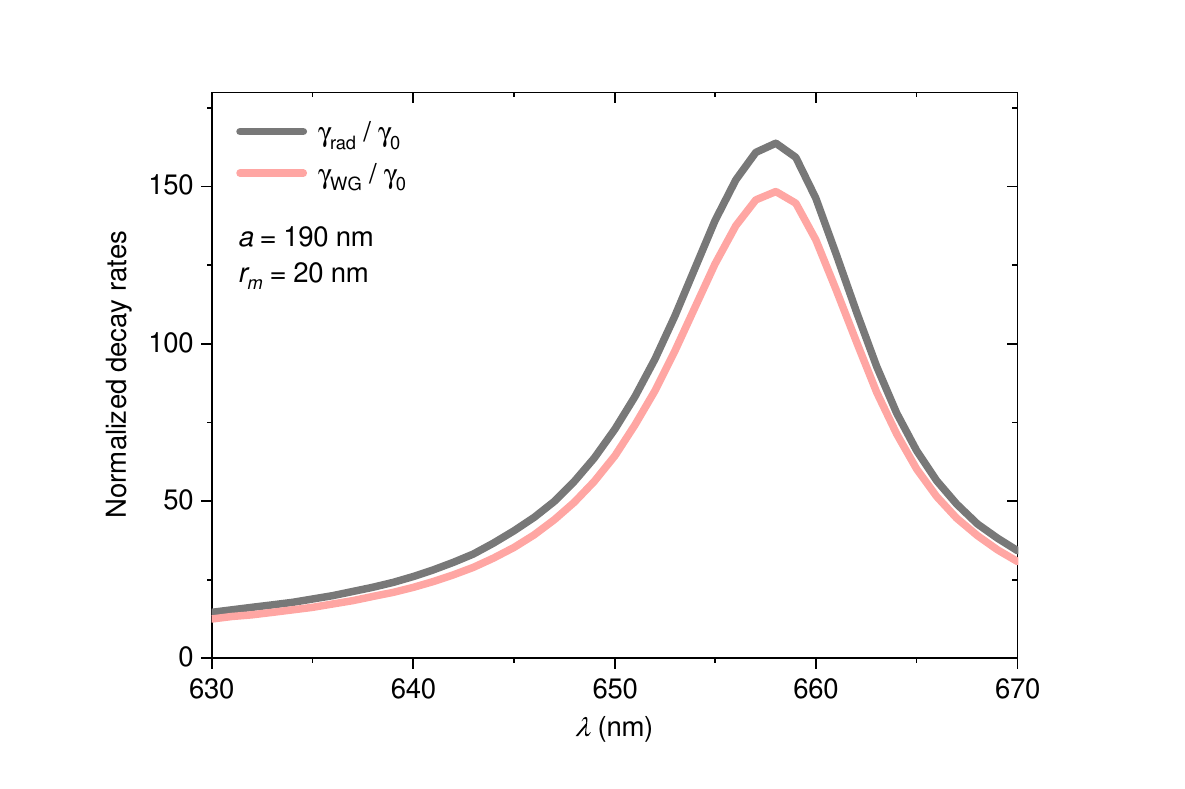}\\
  \caption{Normalized decay rates of the $\sigma_-$ emitter in the coupled PC and AgNP structure for radiative part ($\gamma_{\rm rad}/\gamma_0$) and along the waveguide ($\gamma_{\rm WG}/\gamma_0$). More than $90\%$ of the radiative photons is coupled to the waveguide of the PC.}\label{figS7}
\end{figure}

\clearpage
\begin{figure}
  \centering
  \includegraphics[width=15cm]{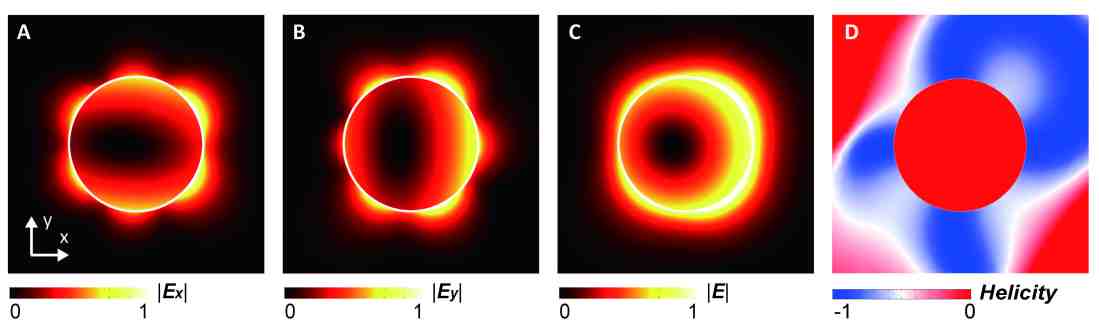}\\
  \caption{The electric field components (A) $|E_x|$ and (B) $|E_y|$, (C) the intensity of the total electric field $|E|$, and (D) the local helicity of $z$ direction at the area of $80 \times 80$ nm$^2$, corresponding to the mode 3 of Fig. 1B in the coupled PC and AgNP structure. The electric fields are normalized by the maximum of $|E|$. By changing the legend of helicity to [-1, 0], the maxima and minima of helicity around the AgNP can be clearly recognized. The parameters are set as $a=190$ nm and $r_m=20$ nm.}\label{figS8}
\end{figure}

\clearpage
\begin{figure}
  \centering
  \includegraphics[width=12cm]{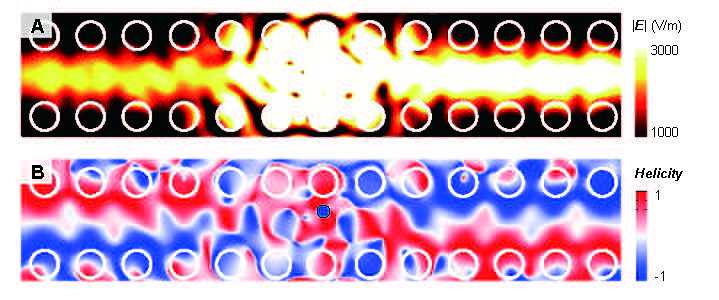}\\
  \caption{(A) Electric field and (B) its helicity of the $\sigma_-$ emitter in the coupled PC and AgNP structure, for $r_m=20$ nm, $y=50$ nm, and $\theta=45^{\circ}$. The photons propagate to two sides along the waveguide of PC, leading to small directionality $D_{\rm R}$. The symmetry of helicity in both propagating channels is not broken compared with that at $\theta=180^{\circ}$ in the main text. }\label{figS9}
\end{figure}

\clearpage
\begin{figure}
  \centering
  \includegraphics[width=13cm]{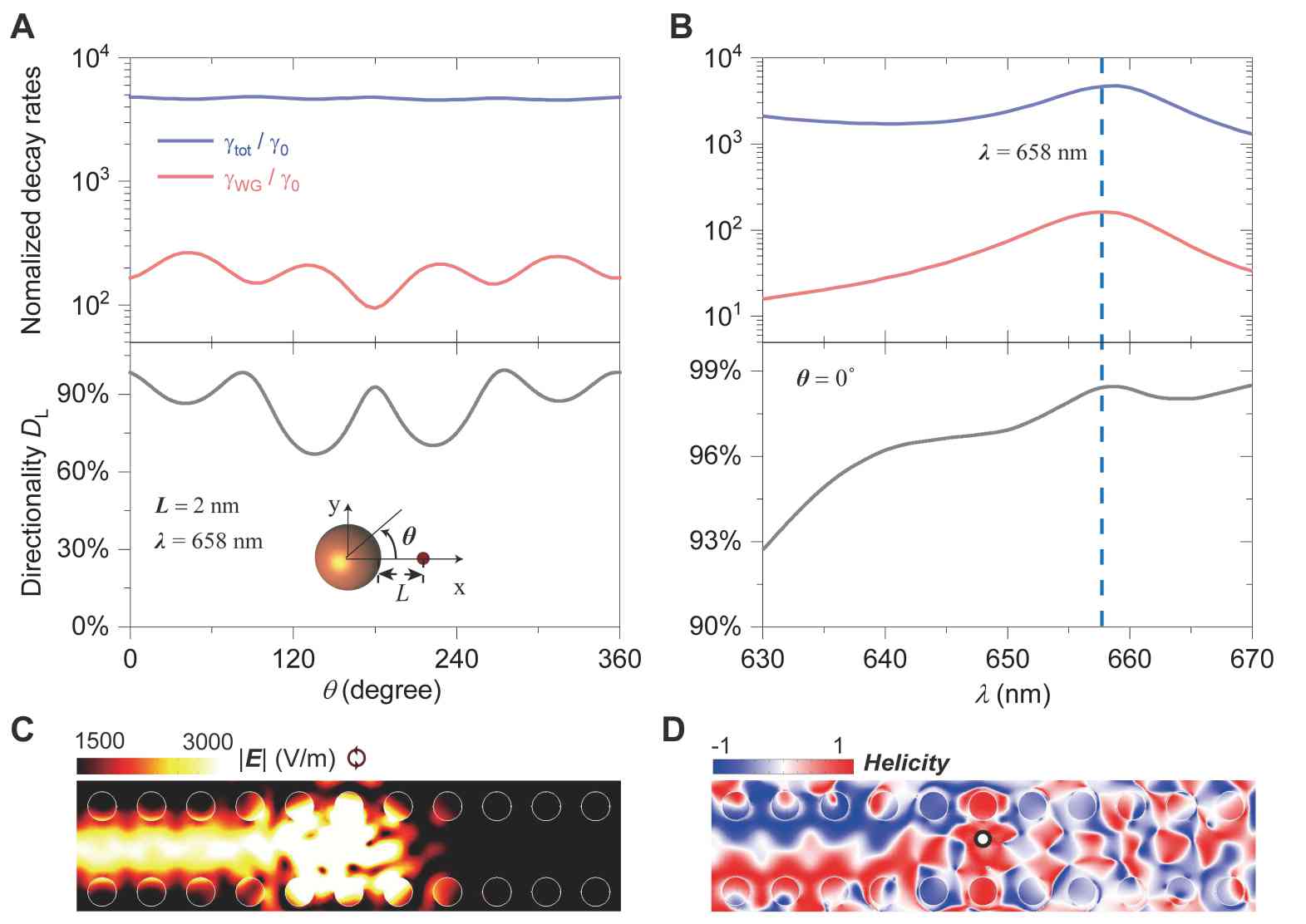}\\
  \caption{Normalized decay rates $\gamma_{\rm tot} / \gamma_0$ and $\gamma_{\rm WG} / \gamma_0$ of a $\sigma_+$ emitter and the directionality $D_{\rm L}$ of the emitted photons as a function of (A) $\theta$ and (B) $\lambda$ for $a=190$ nm, $r_m=20$ nm, and $L=2$ nm. In (A), the maxima of $\gamma_{\rm WG} / \gamma_0$ generally corresponds to minima of $D_{\rm L}$ and vise versa, while $\gamma_{\rm tot} / \gamma_0$ keeps in a high range of $4600 \sim 4800$. $\theta=0^{\circ}$ is chosen in (B). When $\lambda=658$ nm, Total Purcell enhancement reaches $\gamma_{\rm tot} / \gamma_0 = 4700$ and $\gamma_{WG} / \gamma_0 = 148$, where $98.4\%$ of emitted photons propagates to left direction. The distributions of (C) the electric field and (D) its helicity when a $\sigma_+$ emitter excites the quadrupolar mode of the AgNP. From the helicity distribution, the symmetry of the propagation is broken, whose direction is opposite to that of the $\sigma_-$ emitter in Fig. 2D.
}\label{figS10}
\end{figure}

\clearpage
\begin{figure}
  \centering
  \includegraphics[width=13cm]{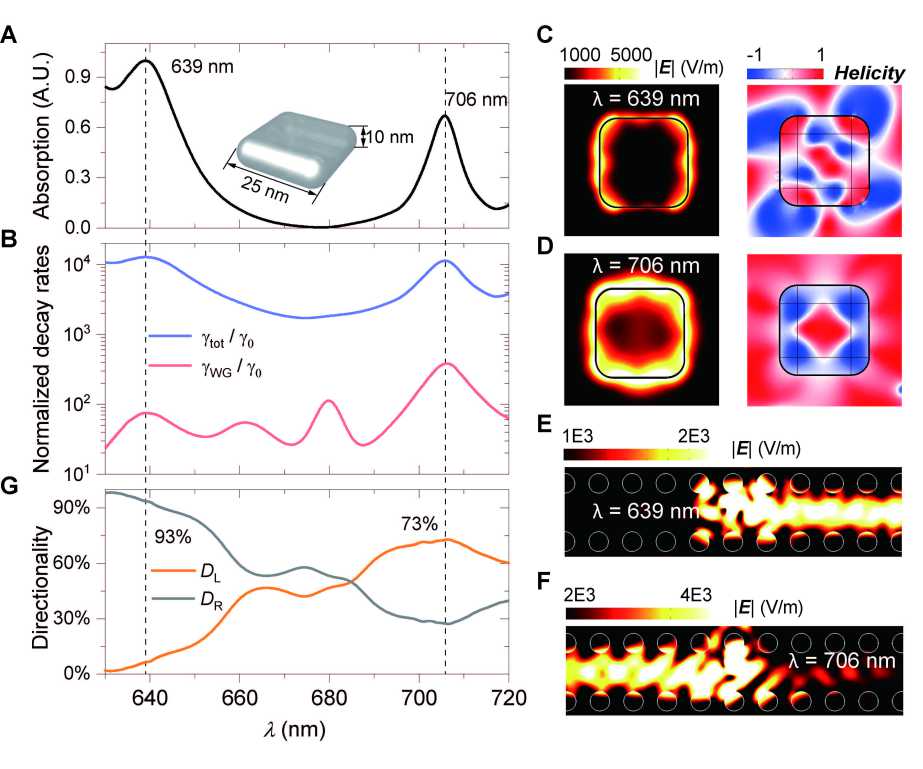}\\
  \caption{(A) Absorption spectra of a silver nano-block with the length of 25 nm and the height of 10 nm embedded in the waveguide of PC for $a=180$ nm. Two adjacent resonance wavelengthes of the block are 639 nm and 706 nm. (B) Normalized decay rate as a function of $\lambda$. Corresponding to two resonant wavelengthes, $\gamma_{\rm tot}/\gamma_0=12719$ and $11428$, and $\gamma_{\rm WG}/\gamma_0=76$ and $389$, respectively.
  (C) The electric field and (D) helicity distributions for the nano-block in the resonance wavelengthes, where the sign of helicity at the corner of the AgNP are opposite. By using this character of the helicity of the two modes, for $\lambda=639$ nm, $93\%$ of the photons emitted from $\sigma_-$ emitter is guided to right direction and $73\%$ to left direction for $\lambda=706$ nm. The electric field (E) for $\lambda=639$ nm and (F) for $\lambda=706$ nm. (G) Directionality $D_{\rm R/L}$ dependent on $\lambda$. Here, the nano-block is placed at $y=50$ nm. }\label{figS11}
\end{figure}

\clearpage
\begin{figure}
  \centering
  \includegraphics[width=12cm]{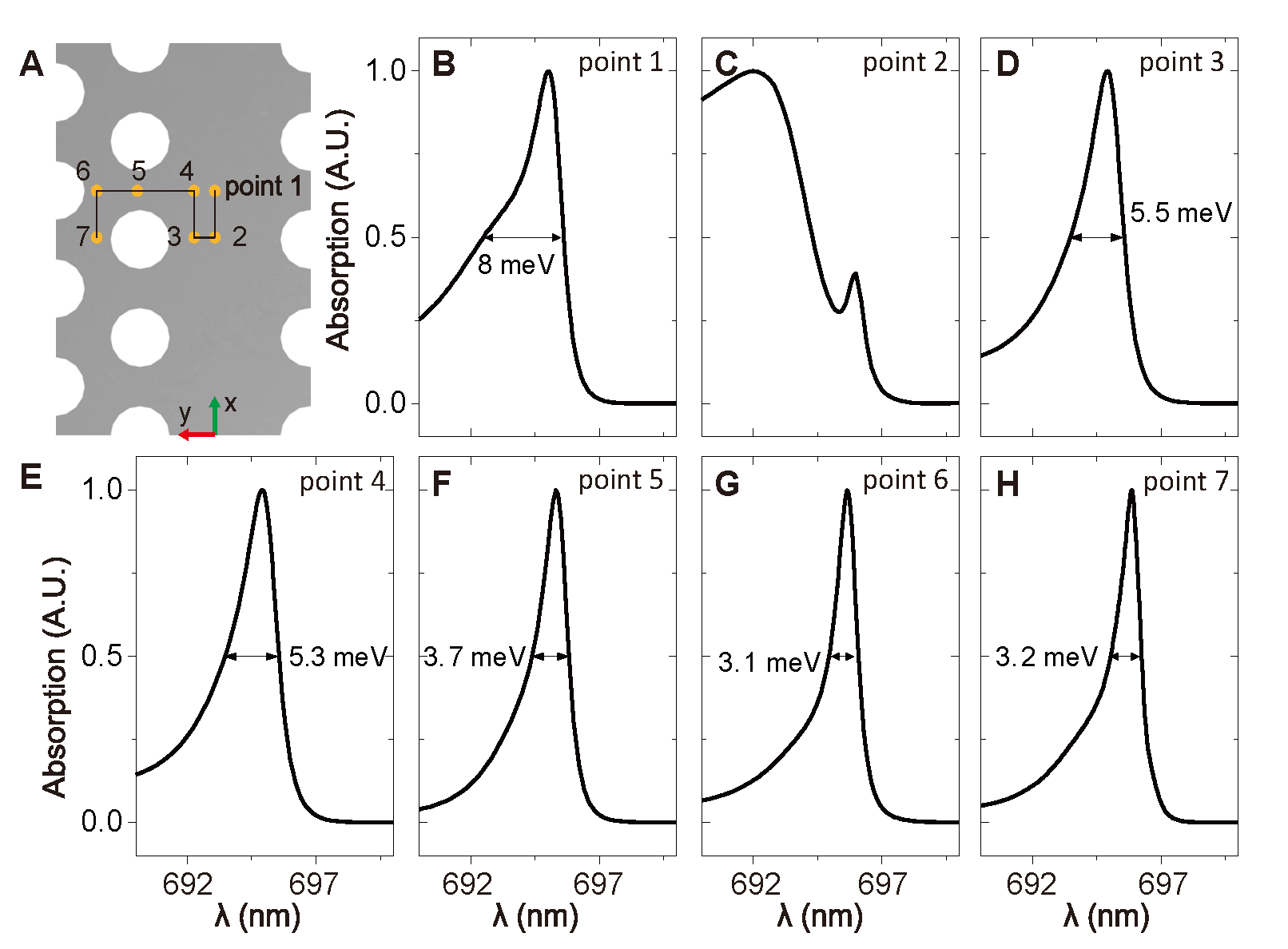}\\
  \caption{(A) Schematic diagram of the AgNP with $r_m=7$ nm located in different positions of the PC. (B to H) The absorption spectra of the band-edge mode for various AgNP positions marked in (A). The linewidth $\kappa$ of the AgNP is optimized by selecting its position. In the main text, point 6 with $\kappa=3.1$ meV is used.}\label{figS12}
\end{figure}

\clearpage
\begin{figure}
  \centering
  \includegraphics[width=15cm]{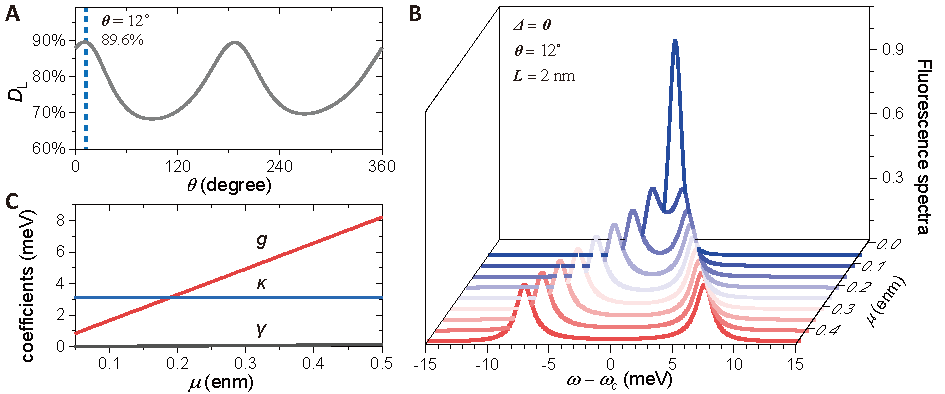}\\
  \caption{(A) Directionality $D_{\rm L}$ of emitted photons dependent on $\theta$. The AgNP with $r_m=7$ nm is placed at point 6 of Fig. S12. Directionality $D_{\rm L}$ reaches the maximum when $\theta=12^{\circ}$ and $190^{\circ}$ with the value of $89.6\%$. (B) Fluorescence spectra of the quantum emitter as a function of $\omega-\omega_c$ with varying dipole moment $\mu$. The parameters $g,~\gamma,~\kappa$ in dependence on $\mu$ are shown in (C). Here, $L=2$ nm and the detunint $\Delta$ between the AgNP and emitter is zero. }\label{figS13}
\end{figure}




\bibliographystyle{Science}


\end{document}